\documentclass[11pt]{article}
\usepackage{amsmath}
\usepackage{amsfonts}
\usepackage{amssymb}
\usepackage[dvips]{graphicx}
\usepackage{color}

\input{tcilatex}
\begin{document}

\title{Scalar $QCD_{4}$ on the null-plane}
\author{R. Casana$^1$\thanks{%
casana@ufma.br}~, B.M. Pimentel$^2$\thanks{%
pimentel@ift.unesp.br}~ and G. E. R. Zambrano$^2$\thanks{%
gramos@ift.unesp.br} \thanks{
On leave of absence from Departamento de F\'{\i}sica, Universidad de Nari%
\~{n}o, San Juan de Pasto, Nari\~{n}o, Colombia} \\
\emph{{\small $^1$ Departamento de F\'{\i}sica, Universidade Federal do Maranh\~{a}o (UFMA),}} \\
\emph{{\small Campus Universit\'{a}rio do Bacanga, CEP 65080-040, S\~{a}o Lu%
\'{\i}s - MA, Brasil.}} \\
\textit{{\small $^2$ Instituto de F\'{\i}sica Te\'orica (IFT/UNESP), UNESP -
S\~ao Paulo State University}} \\
\textit{\small Rua Pamplona 145, CEP 01405-900, S\~ao Paulo, SP, Brazil}}
\date{}
\maketitle

\begin{abstract}
We have studied the null-plane hamiltonian structure of the free
Yang-Mills fields and the scalar chromodynamics ($SQCD_{4}$).
Following the Dirac's procedure for constrained systems we have
performed a detailed analysis of the constraint structure of both
models and we give the generalized Dirac brackets for the physical
variables. In the free Yang-Mills case, using the correspondence
principle in the Dirac's brackets we obtain the same commutators
present in the literature.
\end{abstract}

\section{Introduction}

To quantize the theory on the null-plane, initial conditions on the
hyperplane $x^{+}=cte$ and equal $x^{+}$-commutation relations must
be given and the hamiltonian must describe the time evolution from
an
initial value surface to other parallel surface that intersects the $x^{+}$%
-axis at some later time. Although the prescription has a lot of
similarities with the conventional approach there are significant
differences when we perform the quantization of the theory. Inside
the null-plane framework, the lagrangian which describes a given
field theory is singular and at least second class constraints
appear, these can be eliminated by constructing Dirac's brackets
(DB) and the theory can be quantized, via correspondence principle,
in terms of a reduced number of independent fields, the physical
ones. Thus, the Dirac's method \cite{[N1]} allows built the
null-plane hamiltonian and the canonical commutation relations in
terms of the independent fields of the theory.

The quantization of relativistic field theory at the null plane
time, proposed by Dirac \cite{[N2]}, has found important applications \cite%
{[N3]} in both gauge theories and string theory \cite{[N4]}. It is
interesting to observe that the null-plane quantization of a
non-abelian gauge theory using the null-plane gauge condition,
$A_{-}$, identified the transverse components of the gauge field as
the degrees of freedom of the theory and, therefore, the ghost
fields can be eliminated of the quantum action \cite{[N5]}.

In \cite{[N6]}, Tomboulis has quantized the massless Yang-Mills
field in the null-plane gauge $A_{-}^{a}=0$  and derived the Feynman
rules. However, in \cite{[N7]}, McKeon has shown that the null-plane
quantization of this theory leads a set of second-class constraints
in addition to the usual first-class constraints, characteristics of
the usual instant form quantization, what implies in the
introduction of additional ghost fields in the effective lagrangian.
Moreover, in \cite{[N8]}, Morara, Soldati and McCartor have
quantized the theory in the framework of the standard perturbation
approach and they have explained that the difficulties appearing in
the null-plane gauge are overcome using the gauge $A_{+}^{a}=0$,
such gauge provides a generating functional for the renormalized
Green's functions that takes to the Mandelstan-Leibbrandt's
prescription for the free gluon propagator.

On the other hand, in \cite{[N9]}, Neville and Rohlich have studied
the scalar electrodynamics and have obtained the commutation
relations between free fields from the commutations relations of the
free field operators at unequal times but the commutation relation
representing the interaction was not computed but they affirmed to
be derived solving a quantum constraint. This last commutation
relation was determined in \cite{[N10]}, Casana, Pimentel and
Zambrano have calculated all the commutation relations following a
careful analysis of the constraint structure of the theory and the
results obtained are consistent with the specified in the literature
\cite{[N9]}.

In this paper we will discuss firstly the null-plane structure of
the pure Yang-Mills fields and after its interaction with a scalar
complex field ($SQCD_{4}$) following the Dirac's formalism for
constrained systems. The Hamiltonian analysis follows the spirit
outlined in \cite{[N10]}. The work is organized as follow: In the
section 2, we study the free Yang-Mills field, being its constrained
structure analyzed in detail, thus, we classify the constraints and
the appropriated equations of motion of the dynamical variables are
determined by using the extended hamiltonian. The null-plane gauge
is imposed to transform the set of first class constraints in second
class one and, the Diracs's brackets (DB) among the independent
fields are obtained by choosing appropriate boundary conditions on
the fields. In the section 3, the constraint structure of the scalar
chromodynamics ($SQCD_{4}$) is analyzed, the set of constraints is
classified and the correct equations of motion are checked by using
extended hamiltonian as the generator of temporal evolution. Next,
we invert the second class matrix by imposing appropriated boundary
conditions on the fields and we calculate the DB among the
fundamental dynamical variables. Finally, we give our conclusions
and remarks.

\section{Free Yang-Mills field}

For any semi-simple Lie group with structure constant $f_{bc}^{a}$ the
Yang-Mill lagrangian density is
\begin{equation}
\mathcal{L}=-\frac{1}{4}F_{a}^{\mu \nu }F_{\mu \nu }^{a},  \label{2.3}
\end{equation}%
with $F_{\mu \nu }^{a}=\partial _{\mu }A_{\nu }^{a}-\partial _{\nu }A_{\mu
}^{a}+gf_{bc}^{a}A_{\mu }^{b}A_{\nu }^{c}$, the gauge index $a,b,c$ runs
from $1$ to $n$. Such lagrangian is invariant under the following
infinitesimal gauge transformations%
\begin{equation}
\delta A_{a}^{\mu }\left( x\right) =f_{bc}^{a}\Lambda ^{b}\left( x\right)
A_{c}^{\mu }\left( x\right) +\frac{1}{g}\partial ^{\mu }\Lambda _{a}\left(
x\right) .  \label{2.2}
\end{equation}%
with $\Lambda _{a}\left( x\right) $ a infinitesimal function.

In the present work, we specialize for convenience to the $SU(2)$ gauge
group that only has three generators and $f_{bc}^{a}=\varepsilon _{abc}$,
where $\varepsilon _{abc}$ is the Levi-Civita totally antisymmetric tensor
in three dimensions, thus, we can define everything in such way that we can
forget about raising and lowering group indexes. From (\ref{2.3}) we find
the Euler-Lagrange equations%
\begin{equation}
\left( D_{\nu }\right) ^{ab}F_{b}^{\nu \mu }=0,  \label{2.4}
\end{equation}%
where we have defined the covariant derivative $\left( D_{\nu }\right)
^{ab}\equiv \delta _{b}^{a}\partial _{\nu }-g\varepsilon _{abc}A_{\nu }^{c}$.

\subsection{Structure Constraints and Classification}

In the null-plane dynamics, the canonical conjugate momenta are
\begin{equation}
\pi _{a}^{\mu }\equiv \frac{\partial \mathcal{L}}{\partial \left( \partial
_{+}A_{\mu }^{a}\right) }=-F_{a}^{+\mu }~,  \label{2.5}
\end{equation}%
this equation gives the following set of primary constraints%
\begin{equation}
\phi _{a}\equiv \pi _{a}^{+}\approx 0\qquad ,\qquad \phi _{a}^{k}\equiv \pi
_{a}^{k}-\partial _{-}A_{k}^{a}+\partial _{k}A_{-}^{a}-g\varepsilon
_{abc}A_{-}^{b}A_{k}^{c}\approx 0~.  \label{2.7}
\end{equation}%
and the dynamical relation for $A_{-}^{a}$
\begin{equation}
\pi _{a}^{-}=\partial _{+}A_{-}^{a}-\partial _{-}A_{+}^{a}-g\varepsilon
_{abc}A_{+}^{b}A_{-}^{c}~,  \label{2.6}
\end{equation}

At once, the canonical hamiltonian is given by%
\begin{equation}
H_{C}=\int \!\!d^{3}y\mathcal{H}_{C}=\int \!\!d^{3}y\left\{ \frac{1}{2}%
\left( \pi _{a}^{-}\right) ^{2}+\pi _{a}^{-}\left( D_{-}^{x}\right)
^{ab}A_{+}^{b}+\pi _{a}^{i}\left( D_{i}^{x}\right) ^{ab}A_{+}^{b}+\frac{1}{4}%
\left( F_{ij}^{a}\right) ^{2}\right\} .  \label{2.8}
\end{equation}%
Following the Dirac procedure \cite{[N1]}, we define the primary hamiltonian
adding to the canonical hamiltonian the primary constraints
\begin{equation}
H_{P}=\int \!\!d^{3}y\left\{ \frac{1}{2}\left( \pi _{a}^{-}\right) ^{2}+\pi
_{a}^{-}\left( D_{-}^{x}\right) ^{ab}A_{+}^{b}+\pi _{a}^{i}\left(
D_{i}^{x}\right) ^{ab}A_{+}^{b}+\frac{1}{4}\left( F_{ij}^{a}\right)
^{2}+u^{b}\phi _{b}+\lambda _{l}^{b}\phi _{b}^{l}\right\}  \label{2.9}
\end{equation}%
where $u^{b}$ and $\lambda _{l}^{b}$ are their respective Lagrange
multipliers.

The fundamental Poisson brackets $\left( PB\right) $ among fields are%
\begin{equation}
\left\{ A_{\mu }^{a}(x),\pi _{b}^{\nu }(y)\right\} =\delta _{\mu }^{\nu
}\delta _{b}^{a}\delta ^{3}(x-y).  \label{2.10}
\end{equation}

Requiring that $H_{P}$ is the generator of temporal evolutions, the
consistency condition of the primary constraints, i.e. $\left\{ \phi
,H_{P}\right\} =0$, give us for $\phi _{a}$
\begin{equation}
\left\{ \phi _{a}(x),H_{P}\right\} =\left( D_{-}^{x}\right) ^{ab}\pi
_{b}^{-}+\left( D_{i}^{x}\right) ^{ab}\pi _{b}^{i}\equiv G_{a}(x)\approx 0~,
\label{2.11}
\end{equation}
a genuine secondary constraint, which is the Gauss's law. Also, for
$\phi _{a}^{k}$ we obtain
\begin{equation}
\left\{ \phi _{a}^{k}(x),H_{P}\right\} =\left( D_{k}^{x}\right)
^{ab}F_{+-}^{b}+\left( D_{i}^{x}\right) ^{ab}F_{ik}^{b}-2\left(
D_{-}^{x}\right) ^{ab}\lambda _{k}^{b}\approx 0~,  \label{2.12}
\end{equation}%
a differential equation which allows to compute $\lambda _{k}^{b}~$after
imposition of appropriated boundary conditions. The consistency condition of
the secondary constraint yields%
\begin{equation}
\left\{ G_{a}\left( x\right) ,H_{P}\right\} =g\varepsilon
_{acb}A_{+}^{c}\left( x\right) G_{b}\left( x\right) \approx 0,  \label{2.13}
\end{equation}
the Gauss's law is automatically conserved. Then, there are not more
constraints and the equations (\ref{2.7}) and (\ref{2.11}) give the full set
of constraints.

The set of first class constraints is $\left\{ \pi
_{a}^{+}~,~G_{a}\right\} $ and the set of second class constraints
is $\left\{ \phi _{a}^{k}\right\} $ whose PB's are
\begin{equation}
\left\{ \phi _{a}^{k}(x),\phi _{b}^{l}(y)\right\} =-2\delta _{k}^{l}\left(
D_{-}^{x}\right) ^{ab}\delta ^{3}(x-y)  \label{2.14a}
\end{equation}

\subsection{Equations of motion}

Now we check the equations of motion. The time evolution of the fields is
determined by computing their PB's with the so called extended hamiltonian $%
H_{E}$, which is obtained by adding to the primary hamiltonian all the first
class constraints of the theory:
\begin{equation}
\mathrm{H}_{E}=\int \!\!d^{3}y\left\{ \frac{1}{2}\left( \pi _{b}^{-}\right)
^{2}+\pi _{b}^{-}\left( D_{-}^{y}\right) ^{bc}A_{+}^{c}+\pi _{b}^{i}\left(
D_{i}^{y}\right) ^{bc}A_{+}^{c}+\frac{1}{4}\left( F_{ij}^{b}\right)
^{2}+\lambda _{l}^{b}\phi _{b}^{l}+\mathrm{u}^{b}\phi _{b}+\mathrm{v}
^{b}G_{b}\right\} ,  \label{2.15}
\end{equation}
thus, we have the time evolution of the dynamical variables, i.e,~$\dot{\phi}%
=\left\{ \phi ,H_{E}\right\} $, gives
\begin{eqnarray}
\dot{A}_{+}^{a} &=&\mathrm{u}^{a}  \label{2.16.1} \\
\dot{A}_{-}^{a} &=&\smallskip \pi _{a}^{-}+\left( D_{-}^{x}\right)
^{ac}A_{+}^{c}-\left( D_{-}^{x}\right) ^{ab}\mathrm{v}^{b}  \label{2.16.2} \\
[0.2cm] \dot{A}_{k}^{a} &=&\left( D_{k}^{x}\right) ^{ac}A_{+}^{c}+\lambda
_{k}^{a}-\left( D_{k}^{x}\right) ^{ab}\mathrm{v}^{b}  \label{2.16.3} \\
\dot{\pi}_{a}^{+} &=&G_{a}  \label{2.16.4} \\
\dot{\pi}_{a}^{-} &=&-g\varepsilon _{abc}\pi _{b}^{-}A_{+}^{c}+\left(
D_{l}^{x}\right) ^{ab}\lambda _{l}^{b}-g\varepsilon _{bca}\mathrm{v}^{b}\pi
_{c}^{-}  \label{2.16.5} \\[0.2cm]
\dot{\pi}_{a}^{k} &=&-g\varepsilon _{bca}\pi _{b}^{k}A_{+}^{c}+\left(
D_{j}^{x}\right) ^{ab}F_{kj}^{b}-\smallskip \smallskip \left(
D_{-}^{x}\right) ^{ab}\lambda _{k}^{b}-g\varepsilon _{abc}\pi _{c}^{k}%
\mathrm{v}^{b}~,  \label{2.16.6}
\end{eqnarray}%
if we demand consistence with the Euler-Lagrange equation of motion (\ref%
{2.4}) we must to choose $\mathrm{v}^{b}=0$, however, the multiplier $%
\mathrm{u}^{a}$ remains indeterminate.

The Dirac's algorithm requires of as many gauge conditions as first
class constraints there are, nevertheless these conditions should be
compatible with the Euler-Lagrange equations and together with the
first class set they should form a second class set, in such way
that the Lagrange multipliers, corresponding to the first class set,
are determined. Under such considerations, we choose as the first
gauge condition
\begin{equation}
A_{-}^{a}\approx 0,  \label{2.18}
\end{equation}
whose consistency condition $\dot{A}_{-}^{a}=\left\{A_{-}^{a},H_{E} \right\}
\approx 0$ must be compatible with the dynamical equation (\ref{2.6}) thus
we see that if we choose $\mathrm{v}^{b}=0$ in (\ref{2.16.2}) then the Eq.(%
\ref{2.18}) will hold for all times only if
\begin{equation}
\pi _{a}^{-}+\partial _{-}^{x}A_{+}^{a}\approx 0\ ,  \label{2.19}
\end{equation}%
therefore, the equations (\ref{2.18}) and (\ref{2.19}) constitute
our gauge conditions on the null-plane and they are known as the
null-plane gauge.

\subsection{Dirac Brackets}

The gauge fixing conditions transform the first class set into
second class one, the following stage in the Dirac's procedure is to
transform the second class constraints in strong identities. This
demands an alteration of the canonical brackets (PB) to form a new
brackets set, the Dirac's brackets (DB), with which the second class
constraints are automatically satisfied. Thus, the prescription for
determine the DB implies in calculating the inverse of the second
class matrix, for this purpose, we rename the second class
constraints as
\begin{eqnarray}
\Theta _{1} &\equiv &\pi _{a}^{+}\qquad ,\qquad \Theta _{2}\equiv \left(
D_{-}^{x}\right) ^{ab}\pi _{b}^{-}+\left( D_{i}^{x}\right) ^{ab}\pi _{b}^{i}
\notag \\[0.2cm]
\Theta _{3} &\equiv &A_{-}^{a}\qquad ,\qquad \Theta _{4}\equiv \pi
_{a}^{-}+\partial _{-}^{x}A_{+}^{a}~  \label{2.20} \\[0.2cm]
\Theta _{5} &\equiv &\pi _{a}^{k}-\partial _{-}^{x}A_{k}^{a}+\partial
_{k}^{x}A_{-}^{a}-g\varepsilon _{abc}A_{-}^{b}A_{k}^{c},  \notag
\end{eqnarray}%
and we define the elements of the second class matrix as $F_{ab}\left(
x,y\right) \equiv \left\{ \mathbf{\medskip }\Theta _{a}\left( x\right)
,\Theta _{b}\left( y\right) \right\} $. With these considerations, the
Dirac's bracket of two dynamical variables, $\mathbf{A}_{a}\left( x\right) $
and $\mathbf{B}_{b}\left( y\right) $, is then defined as
\begin{equation}
\left\{ \!\frac{{}}{{}}\mathbf{A}_{a}\left( x\right) ,\mathbf{B}_{b}\left(
y\right) \right\} _{D}=\left\{ \!\frac{{}}{{}}\mathbf{A}_{a}\left( x\right) ,%
\mathbf{B}_{b}\left( y\right) \right\} -\int d^{3}ud^{3}v\left\{ \!\frac{{}}{%
{}}\mathbf{A}_{a}\left( x\right) ,\Theta _{c}\left( u\right) \right\} \left(
F^{-1}\right) ^{cd}\left( u,v\right) \left\{ \! \frac{{}}{{}}\Theta
_{d}\left( v\right) ,\mathbf{B}_{b}\left( y\right) \right\} ,  \label{2.21}
\end{equation}%
where $F^{-1}$ is the inverse of the constraint matrix.

The explicit evaluation of $F^{-1}$ involve the determination of an
arbitrary function of the variables $x^{+}$ and $x^{\perp }$ \cite{[N10]}
which can be fixed by considering appropriate boundary conditions \cite%
{[N11]} on the fields $A_{\mu }^{a}$ eliminating the ambiguity in
the definition of the inverse of the operator $\partial _{-}$
related to their zero modes that give origin to hidden subset of
first class constraints which generate improper gauge
transformations \cite{[N13]} what is characteristic of the
null-plane constraint structure \cite{[N12]}. Thus, from
(\ref{2.21}) we obtain the DB among the independent variables of the
theory%
\begin{eqnarray}
\left\{ \mathbf{\medskip }A_{k}^{a}\left( x\right) ,A_{l}^{b}\left( y\right)
\right\} _{D} &=&-\frac{1}{4}\delta _{b}^{a}\delta _{k}^{l}\epsilon \left( \
x-y\right) \delta ^{2}\left( x^{\intercal }-y^{\intercal }\right)  \notag \\
&&  \label{2.22} \\[-0.3cm]
\left\{ \mathbf{\medskip }A_{k}^{a}\left( x\right) ,A_{+}^{b}\left( y\right)
\right\} _{D} &=&\frac{1}{4}\left\vert \ x-y\right\vert \left(
D_{k}^{x}\right) ^{ab}\delta ^{2}\left( x^{\intercal }-y^{\intercal }\right)
.  \notag
\end{eqnarray}%
At once, via the correspondence principle we obtain the commutators among
the fields%
\begin{eqnarray}
\left[ \mathbf{\medskip }A_{k}^{a}\left( x\right) ,A_{l}^{b}\left( y\right) %
\right] &=&-\frac{i}{4}\delta _{b}^{a}\delta _{k}^{l}\epsilon \left( \
x-y\right) \delta ^{2}\left( x^{\intercal }-y^{\intercal }\right) ,
\label{2.23} \\[0.1cm]
\left[ \mathbf{\medskip }A_{k}^{a}\left( x\right) ,A_{+}^{b}\left( y\right) %
\right] &=&\frac{i}{4}\left\vert \ x-y\right\vert \left( D_{k}^{x}\right)
^{ab}\delta ^{2}\left( x^{\intercal }-y^{\intercal }\right) .  \label{2.24}
\end{eqnarray}%
The first relationship is exactly that obtained by Tomboulis \cite{[N6]}and
starting from it is possible to calculate the other two expressions
determined by him, meanwhile the equation (\ref{2.24}) is our contribution
to the commutation relations.

\section{Scalar chromodynamics $SQCD_{4}$}

The model describing the interaction of Yang-Mills and complex scalar field
is given the following lagrangian density
\begin{equation}
\mathcal{L}=\eta ^{\mu \nu }\left( D_{\mu }\right) ^{ab}\Phi _{b}^{\dagger
}\left( D_{\nu }\right) ^{ac}\Phi _{c}-m^{2}\Phi _{a}^{\dagger }\Phi _{a}-%
\frac{1}{4}F_{a}^{\mu \nu }F_{\mu \nu }^{a},  \label{3.1}
\end{equation}%
where the field strength $F_{\mu \nu }^{a}$ and the covariant derivative $%
\left( D_{\mu }\right) ^{ab}$ are defined in the $SU(2)$ adjoint
representation by
\begin{equation}
F_{\mu \nu }^{a}=\partial _{\mu }A_{\nu }^{a}-\partial _{\nu }A_{\mu
}^{a}+g\varepsilon _{abc}A_{\mu }^{b}A_{\nu }^{c}\qquad ,\qquad \left(
D_{\mu }\right) ^{ab}\equiv \delta _{a}^{b}\partial _{\mu }-g\varepsilon
_{abc}A_{\mu }^{c}~,  \label{3.2}
\end{equation}%
respectively. $\Phi _{c}$ is the complex scalar field which has
three components in an internal space and the gauge transformation
are rotations in this space what gives a conserved vector quantity
named \textit{isospin}. The field equations are given for
\begin{eqnarray}
\left( D_{\nu }\right) ^{ab}F_{b}^{\nu \mu } &=&J_{a}^{\mu }  \notag \\
\left( D_{\mu }\right) ^{ab}\left( D^{\mu }\right) ^{bc}\Phi _{c}+m^{2}\Phi
_{a} &=&0  \label{3.3} \\
\left( D^{\mu }\right) ^{ab}\left( D_{\mu }\right) ^{bc}\Phi _{c}^{\dagger
}+m^{2}\Phi _{a}^{\dagger } &=&0,  \notag
\end{eqnarray}%
where $J_{h}^{\beta }$ is the current density defined by%
\begin{equation}
J_{a}^{\mu }\equiv g\varepsilon _{abc}\left\{ \smallskip \left[ \strut
\smallskip \left( D^{\mu }\right) ^{bd}\Phi _{d}^{\dagger }\right] \Phi
_{c}+\Phi _{c}^{\dagger }\left[ \strut \smallskip \left( D^{\mu }\right)
^{bd}\Phi _{d}\right] \right\} .  \label{3.4}
\end{equation}

\subsection{Structure Constraints and Classification}

The canonical conjugate momenta of the gauge field is
\begin{equation}
\pi _{a}^{\mu }\equiv \frac{\partial \mathcal{L}}{\partial \left( \partial
_{+}A_{\mu }^{a}\right) }=-F_{a}^{+\mu }~,  \label{3.5}
\end{equation}%
and for the fields $\Phi _{a}~,~\Phi _{a}^{\dagger }$ are%
\begin{equation}
\Pi _{a}^{\dagger }\equiv \frac{\partial \mathcal{L}}{\partial \left(
\partial _{+}\Phi _{a}\right) }=\left( D_{-}\right) ^{ab}\Phi _{b}^{\dagger
}\quad ,\quad \Pi _{a}\equiv \frac{\partial \mathcal{L}}{\partial \left(
\partial _{+}\Phi _{a}^{\dagger }\right) }=\left( D_{-}\right) ^{ab}\Phi _{b}
\label{3.5.a}
\end{equation}%
respectively.

From (\ref{3.5}) we get one dynamical relation for $A_{-}^{a}$
\begin{equation}
\pi _{a}^{-}=\partial _{+}A_{-}^{a}-\partial _{-}A_{+}^{a}-g\varepsilon
_{abc}A_{+}^{b}A_{-}^{c},  \label{3.6}
\end{equation}%
and the following set of primary constraints for the gauge sector%
\begin{equation}
\phi _{a}\equiv \pi _{a}^{+}\approx 0\qquad ,\qquad \phi _{a}^{k}\equiv \pi
_{a}^{k}-\partial _{-}A_{k}^{a}+\partial _{k}A_{-}^{a}-g\varepsilon
_{abc}A_{-}^{b}A_{k}^{c}\approx 0,  \label{3.7}
\end{equation}%
and from (\ref{3.5.a}) we obtain a set of primary constraints of the scalar
sector%
\begin{equation}
\Theta _{a}\equiv \Pi _{a}-\left( D_{-}\right) ^{ab}\Phi _{b}\approx 0\qquad
,\qquad \Theta _{a}^{\dagger }\equiv \Pi _{a}^{\dagger }-\left( D_{-}\right)
^{ab}\Phi _{b}^{\dagger }\approx 0.  \label{3.8}
\end{equation}

The canonical hamiltonian is
\begin{eqnarray}
H_{C} &=&\int \!\!d^{3}y\smallskip \left\{ \frac{1}{2}\left( \pi
_{b}^{-}\right) ^{2}+\pi _{b}^{-}\left( D_{-}\right) ^{bc}A_{+}^{c}+\pi
_{b}^{k}\left( D_{k}\right) ^{bc}A_{+}^{c}+J_{c}^{+}A_{+}^{c}\right\} \
\notag \\[-0.3cm]
&& \\
&&+\int \!\!d^{3}y\smallskip \left\{ \left[ \left( D_{k}^{x}\right)
^{ab}\Phi _{b}^{\dagger }\right] \left[ \left( D_{k}^{x}\right) ^{ad}\Phi
_{d}\right] +m^{2}\Phi _{b}\Phi _{b}+\frac{1}{4}\left( F_{jk}^{b}\right)
^{2}\right\}  \notag
\end{eqnarray}%
and the primary hamiltonian is%
\begin{equation}
H_{P}=H_{C}+\int \!\!d^{3}y\smallskip \left\{ \frac{{}}{{}}\!u^{b}\phi
_{b}+\lambda _{k}^{b}\phi _{b}^{k}+U_{b}^{\dagger }\Theta _{b}\smallskip
+\Theta _{b}^{\dagger }U_{b}\right\} ,  \label{3.10}
\end{equation}%
where $u^{b}$ and $\lambda _{k}^{b}$ are the Lagrange multipliers associated
to the vector constraints and $U_{b}^{\dagger }$ and $U_{b}$ are the
multipliers associated with the scalar ones.

The fundamental Poisson brackets are%
\begin{equation}
\left\{ A_{\mu }^{a}\left( x\right) ,\pi _{b}^{\nu }\left( y\right) \right\}
=\delta _{\mu }^{\nu }\delta _{b}^{a}\delta ^{3}\left( x-y\right)
\end{equation}%
\begin{equation}
\left\{ \Phi _{a}\left( x\right) ,\Pi _{b}^{\dagger }\left( y\right)
\right\} =\delta _{a}^{b}\delta ^{3}\left( x-y\right) \quad ,\quad \left\{
\Phi _{a}^{\dagger }\left( x\right) ,\Pi _{b}\left( y\right) \right\}
=\delta _{a}^{b}\delta ^{3}\left( x-y\right) ,  \label{3.11}
\end{equation}%
and the non null PB's among the primary constraints%
\begin{eqnarray}
\left\{ \phi _{a}^{k}(x),\phi _{b}^{l}(y)\right\} &=&-2\delta _{k}^{l}\left(
D_{-}^{x}\right) ^{ab}\delta ^{3}\left( x-y\right) ~,  \notag \\[-0.4cm]
&& \\
\left\{ \Theta _{a}\left( x\right) ,\Theta _{b}^{\dagger }\left( y\right)
\right\} &=&-2\left( D_{-}^{x}\right) ^{ab}\delta ^{3}\left( x-y\right) ~.
\notag
\end{eqnarray}

Following the Dirac's procedure, we compute the consistence condition of
every primary constraint. Thus, the consistence condition of the scalar
constraints yields:%
\begin{eqnarray}
\dot{\Theta}_{a} &=&-g\varepsilon _{abc}\pi _{b}^{-}\Phi _{c}-2g\varepsilon
_{bcd}\left( D_{-}\right) ^{ab}\left[ \Phi _{d}A_{+}^{c}\right] +\left(
D_{k}\right) ^{ab}\left( D_{k}\right) ^{bc}\Phi _{c}-\smallskip m^{2}\Phi
_{a}-2\left( D_{-}\right) ^{ab}U_{b}  \notag \\[-0.4cm]
&&  \label{3.13} \\
\dot{\Theta}_{a}^{\dagger } &=&-g\varepsilon _{abc}\pi _{b}^{-}\Phi
_{c}^{\dagger }-2g\varepsilon _{bcd}\left( D_{-}\right) ^{ab}\smallskip %
\left[ \Phi _{d}^{\dagger }A_{+}^{c}\right] +\smallskip \left( D_{k}\right)
^{ab}\left( D_{k}\right) ^{bc}\Phi _{c}^{\dagger }-m^{2}\Phi _{a}^{\dagger
}\smallskip -2\smallskip \left( D_{-}\right) ^{ab}U_{b}^{\dagger }~,  \notag
\end{eqnarray}%
these relations allow to determine the multipliers $U_{b}$ and $%
U_{b}^{\dagger }$, respectively. In this way, there are not more constraints
associated with the scalar sector.

In the gauge sector, the consistency condition of $\phi _{a}^{k}$ provides%
\begin{equation}
\dot{\phi}_{a}^{k}=\left( D_{k}\right) ^{ab}\smallskip \pi _{b}^{-}+\left(
D_{j}\right) ^{ab}F_{jk}^{b}-J_{a}^{k}-2\left( D_{-}\right) ^{ab}\lambda
_{k}^{b}\approx 0  \label{3.14}
\end{equation}%
that is an equation which determines the multiplier $\lambda
_{k}^{b}$. Finally, the consistence condition of $\pi _{a}^{+}$
contributes with a
secondary constraint%
\begin{equation}
\dot{\phi}_{a}=\left( D_{-}\right) ^{ab}\pi _{b}^{-}+\left( D_{i}\right)
^{ab}\pi _{b}^{i}-J_{a}^{+}\equiv G_{a}\approx 0,  \label{3.15}
\end{equation}%
which is the Gauss's law for the scalar chromodynamics. After a
laborious work, it is possible to verify that no more further
constraints are generated from the consistence condition of the
Gauss' law because it is
automatically conserved%
\begin{equation}
\dot{G}_{a}=g\varepsilon _{abc}\left[ \Phi _{c}^{\dagger }\dot{\Theta}%
_{b}+\Phi _{c}\dot{\Theta}_{b}^{\dagger }\smallskip \right] \approx 0.
\label{3.16}
\end{equation}%
Therefore, the equation (\ref{3.7}), (\ref{3.8}) and (\ref{3.15}) constitute
the full set of constraints of the theory.

The non null PB's among the constraints of the theory are%
\begin{eqnarray}
\left\{ \phi _{a}^{k}(x),\phi _{b}^{l}(y)\right\} &=&-2\delta _{k}^{l}\left(
D_{-}^{x}\right) ^{ab}\delta ^{3}\left( x-y\right) ~~,~\ \ \ \   \notag
\label{3.17} \\
\left\{ \Theta _{a}^{\dagger }\left( x\right) ,\Theta _{b}\left( y\right)
\right\} &=&-2\left( D_{-}^{x}\right) ^{ab}\delta ^{3}\left( x-y\right) ~~,
\notag \\[-0.5cm]
&& \\
\left\{ \smallskip G_{a}\left( x\right) ,\Theta _{b}^{\dagger }\left(
y\right) \right\} &=&-2g\varepsilon _{acf}\Phi _{f}^{\dagger }\left(
x\right) \smallskip \smallskip \left( D_{-}^{x}\right) ^{cb}\delta
^{3}\left( x-y\right) ~~,  \notag \\
\left\{ \smallskip G_{a}\left( x\right) ,\Theta _{b}\left( y\right) \right\}
&=&-2g\varepsilon _{acf}\Phi _{f}\left( x\right) \left( D_{-}^{x}\right)
^{cb}\delta ^{3}\left( x-y\right) ~~,\   \notag
\end{eqnarray}%
thus, it is easy to note that $\pi _{a}^{+}$ is vanishing PB with
all the other constraints, therefore, it is a first class
constraint. The remaining set, $\left\{ \phi _{a}^{k},~\Theta
_{a},~\Theta _{a}^{\dagger },~G_{a}\right\} $, is apparently a
second class set, however, it is possible to show that their
constraint matrix is singular and its zero mode eigenvector provides
a linear combination of constraints which is a first
class constraint \cite{[N10]}. Such second first class constraint is%
\begin{equation}
\Sigma _{a}\equiv G_{a}-g\varepsilon _{abc}\left[ \Phi _{c}^{\dagger }\Theta
_{b}+\Phi _{c}\Theta _{b}^{\dagger }\smallskip \right] .  \label{3.18}
\end{equation}

Then, the first class constraints set is
\begin{equation}
\phi _{a}=\pi _{a}^{+}\qquad ,\qquad \Sigma _{a}=G_{a}-g\varepsilon _{abc}
\left[ \Phi _{c}^{\dagger }\Theta _{b}+\Phi _{c}\Theta _{b}^{\dagger
}\smallskip \right] .  \label{3.20}
\end{equation}%
it is the maximal number of first class constraints and, the second class
set is%
\begin{eqnarray}
\phi _{a}^{k} &=&\pi _{a}^{k}-\partial _{-}A_{k}^{a}+\partial
_{k}A_{-}^{a}-g\varepsilon _{abc}A_{-}^{b}A_{k}^{c}~,  \notag \\
\Theta _{a} &=&\Pi _{a}-\left( D_{-}\right) ^{ab}\Phi _{b}~,  \label{3.19} \\
\Theta _{a}^{\dagger } &=&\Pi _{a}^{\dagger }-\left( D_{-}\right) ^{ab}\Phi
_{b}^{\dagger }~.  \notag
\end{eqnarray}

\subsection{Equations of Motion}

At this point we need to check that we have the correct equation of motion.
The time evolution of the fields is determined by computing their PB with
the extended hamiltonian which is defined as
\begin{equation}
H_{E}=H_{C}+\int d^{3}y\left\{ \mathrm{u}^{b}\phi _{b}+\mathrm{\lambda }%
_{l}^{b}\phi _{b}^{l}+\mathrm{U}_{b}^{\dagger }\Theta _{b}\smallskip +\Theta
_{b}^{\dagger }\mathrm{U}_{b}+\mathrm{w}_{b}\Sigma _{b}\right\}  \label{3.21}
\end{equation}%
Thus, the time evolution of the gauge field yields%
\begin{eqnarray}
\dot{A}_{+}^{a} &=&\mathrm{u}^{a}  \notag \\
\dot{A}_{-}^{a} &=&\smallskip \pi _{a}^{-}+\left( D_{-}\right)
^{ac}A_{+}^{c}-\left( D_{-}\right) ^{ab}\mathrm{w}_{b}  \notag \\
\dot{A}_{k}^{a} &=&\left( D_{k}\right) ^{ab}A_{+}^{b}+\mathrm{\lambda }%
_{k}^{a}-\left( D_{k}\right) ^{ab}\mathrm{w}_{b}  \notag \\
\dot{\pi}_{a}^{+} &=&G_{a}  \notag \\
\dot{\pi}_{a}^{-} &=&g\varepsilon _{abc}\pi
_{b}^{-}A_{+}^{c}+2g^{2}\smallskip A_{+}^{a}\Phi _{b}^{\dagger }\Phi
_{b}-g^{2}\smallskip A_{+}^{b}\Phi _{a}^{\dagger }\Phi _{b}-g^{2}\smallskip
A_{+}^{b}\Phi _{b}^{\dagger }\Phi _{a}  \label{3.22.1} \\
&&+\left( D_{k}\right) ^{ab}\mathrm{\lambda }_{k}^{b}-g\varepsilon _{abc}%
\mathrm{U}_{b}^{\dagger }\Phi _{c}-g\varepsilon _{abc}\Phi _{c}^{\dagger }%
\mathrm{U}_{b}+g\varepsilon _{bca}\mathrm{w}_{b}\pi _{c}^{-}  \notag \\%
[0.2cm]
\dot{\pi}_{a}^{k} &=&g\varepsilon _{abc}\pi
_{b}^{k}A_{+}^{c}-J_{a}^{k}+\smallskip \left( D_{j}\right)
^{ab}F_{jk}^{b}-\left( D_{-}\right) ^{ab}\mathrm{\lambda }%
_{k}^{b}+g\varepsilon _{abc}\mathrm{w}_{b}\pi _{c}^{k}  \notag
\end{eqnarray}%
and for the scalar fields the dynamics is given for%
\begin{eqnarray}
\dot{\Phi}_{a} &=&\mathrm{U}_{a}+g\varepsilon _{abc}\mathrm{w}_{b}\Phi _{c}~,
\notag \\
\dot{\Phi}_{a}^{\dagger } &=&\mathrm{U}_{a}^{\dagger }+g\varepsilon _{abc}%
\mathrm{w}_{b}\Phi _{c}^{\dagger }~,  \notag \\
\dot{\smallskip \smallskip \Pi }_{a} &=&g\varepsilon _{cde}\left(
D_{-}\right) ^{ad}\smallskip \smallskip \left[ \smallskip A_{+}^{c}\Phi _{e}%
\right] -g\varepsilon _{abc}A_{+}^{b}\left[ \strut \smallskip \left(
D_{-}\right) ^{cd}\Phi _{d}\right] +\left( D_{k}\right) ^{ab}\left[ \left(
D_{k}\right) ^{bc}\Phi _{c}\right]  \label{3.23} \\
&&-m^{2}\Phi _{a}-\left( D_{-}\right) ^{ab}\mathrm{U}_{b}-g\varepsilon _{abc}%
\mathrm{w}_{b}\Phi _{c}^{\dagger }~,  \notag \\
\dot{\smallskip \smallskip \Pi }_{a}^{\dagger } &=&-g\varepsilon
_{abc}A_{+}^{b}\left[ \strut \smallskip \left( D_{-}\right) ^{cd}\Phi
_{d}^{\dagger }\right] +g\varepsilon _{bcd}\left( D_{-}\right) ^{ac}\left[
\smallskip A_{+}^{b}\Phi _{d}^{\dagger }\right] \strut \smallskip +\left(
D_{k}\right) ^{ab}\left[ \left( D_{k}\right) ^{bc}\Phi _{c}^{\dagger }\right]
\notag \\
&&-m^{2}\Phi _{a}^{\dagger }-\left( D_{-}\right) ^{ab}\mathrm{U}%
_{b}^{\dagger }+g\varepsilon _{abd}\mathrm{w}_{b}\Pi _{d}^{\dagger }~,
\notag
\end{eqnarray}%
We can note that the set of equation (\ref{3.22.1}) and (\ref{3.23})
only will be compatible with the Euler-Lagrange equations
(\ref{3.3}) if we set $\mathrm{w}_{b}=0$ however the multiplier
$\mathrm{u}^{a}$ still remains undetermined in this way the Dirac's
formalism tell us to impose one set of gauge conditions, one for
every first class constraint. The gauge conditions are chosen in
such a way that they are compatible with the Euler-Lagrange
equations of motion, thus one such set is the null-plane gauge
conditions is given by relations (\ref{2.18}) and (\ref{2.19}).

\subsection{Dirac Brackets}

We have the following set of second class constraints:%
\begin{eqnarray}
\Psi _{1} &\equiv &\pi _{a}^{+}\quad ,\quad \Psi _{2}\equiv
G_{a}-g\varepsilon _{abc}\left( \Phi _{c}^{\dagger }\Theta _{b}+\Phi
_{c}\Theta _{b}^{\dagger }\smallskip \right)   \notag \\[0.2cm]
\Psi _{3} &\equiv &A_{-}^{a}\quad ,\quad \Psi _{4}\equiv \pi
_{a}^{-}+\partial _{-}A_{+}^{a}  \notag \\[-0.4cm]
&&  \label{3.26} \\
\Psi _{5} &\equiv &\pi _{a}^{k}-\partial _{-}A_{k}^{a}+\partial
_{k}A_{-}^{a}-g\varepsilon _{abc}A_{k}^{c}A_{-}^{b}  \notag \\[0.2cm]
\Psi _{6} &\equiv &\Pi _{a}-\left( D_{-}\right) ^{ab}\Phi _{b}\quad ,\quad
\Psi _{7}\equiv \Pi _{a}^{\dagger }-\left( D_{-}\right) ^{ab}\Phi
_{b}^{\dagger },  \notag
\end{eqnarray}%
with these, we define the following constraint matrix $M_{ab}\left(
x,y\right) \equiv \left\{ \mathbf{\medskip }\Psi _{a}\left( x\right) ,\Psi
_{b}\left( y\right) \right\} $, from where the DB for the dynamical
variables are determined via evaluation of the inverse of this matrix.Then,
by considering appropriate boundary conditions on the fields, \cite%
{[N10],[N11]}, a unique inverse of the constraint matrix is obtained and
after a laborious work we obtain the DB for the independent dynamical
variables of scalar chromodynamics:%
\begin{eqnarray}
\left\{ \smallskip A_{k}^{a}\left( x\right) ,A_{l}^{b}\left( y\right)
\right\} _{D} &=&-\frac{1}{4}\delta _{b}^{a}\delta _{k}^{l}\epsilon \left( \
x-y\right) \delta ^{2}\left( x^{\intercal }-y^{\intercal }\right)   \notag \\
\left\{ \smallskip \Phi _{a}\left( x\right) ,\Phi _{b}^{\dagger }\left(
y\right) \right\} _{D} &=&-\frac{1}{4}\delta _{a}^{b}\epsilon \left(
x-y\right) \delta ^{2}\left( x^{\intercal }-y^{\intercal }\right)
\label{3.27} \\
\left\{ \smallskip \Phi _{a}\left( x\right) ,A_{+}^{b}\left( y\right)
\right\} _{D} &=&\frac{g}{2}\varepsilon _{abc}\delta ^{2}\left( x^{\intercal
}-y^{\intercal }\right) \left\{ \smallskip \Phi _{c}\left( x\right)
\left\vert \ x-y\right\vert -\frac{1}{4}\int dv~\Phi _{c}\left( v\right)
\epsilon \left( x-v\right) \epsilon \left( \ v-y\right) \right\}   \notag \\
\left\{ \mathbf{\medskip }\Phi _{a}^{\dagger }\left( x\right)
,A_{+}^{b}\left( y\right) \right\} _{D} &=&\frac{g}{2}\varepsilon
_{abc}\delta ^{2}\left( x^{\intercal }-y^{\intercal }\right) \left\{ \Phi
_{c}^{\dagger }\left( x\right) \left\vert \ x-y\right\vert -\frac{1}{4}\int
dv~\Phi _{c}^{\dagger }\left( v\right) \epsilon \left( x-v\right) \epsilon
\left( \ v-y\right) \right\}   \notag
\end{eqnarray}

\section{Remarks and conclusions}

In this work we have studied the null plane Hamiltonian structure of
the free Yang-Mills field and its interaction with a complex scalar
that we named as scalar chromodynamics ($SQCD_{4}$).

Performing a careful analysis of the constraint structure of
Yang-Mills field, we have determined in addition of the usual first
class constraints set a second class ones set, which is a
characteristic of the null-plane dynamics \cite{[N10]}. The
imposition of appropriated boundary conditions on the fields fixes
the hidden subset of first class constraints \cite{[N12]} and
eliminates the ambiguity on the operator $\partial_{-}$, that allows
to get a unique inverse for the second class constraint matrix
\cite{[N10]}. The Dirac's brackets of the theory are quantized via
correspondence principle; the commutators obtained are the same
derived by Tomboulis \cite{[N6]}.

The scalar chromodynamics $SQCD_{4}$ hamiltonian analysis has shown
further of the free Yang-Mill structure, a contribution of the
scalar sector  with an additional constraints set. However,  as a
consequence  of the constraint associated with the scalar part, one
of the first class constraints  is a linear combination of the
Gauss' law with the scalar constraints, in a similar way to the
scalar electrodynamics case \cite{[N10]}, such first class
constraint is given by the zero mode eigenvector of the constraint
matrix. Finally, choosing the null-plane gauge condition, which
transforms first class constraints in second class ones and imposing
appropriated boundary conditions  on the fields to get a unique
inverse  of the second class constraints matrix and following the
Dirac' procedure we obtain the Dirac' brackets of the canonical
variables of the theory. Our results are consistent with those
reported in the literature \cite{[N9],[N10]} when the abelian case
is considered.

As the null-plane hamiltonian structure is well-defined, the
null-plane quantization, of the models reported here and
\cite{[N10]}, via the path-integral formalism are now in advanced
and whose result will be reported elsewhere.

\subsection*{Acknowledgements}

RC thanks to CNPq for partial support, BMP thanks CNPq for partial support
and GERZ thanks CNPq (grant 142695/2005-0) for full support.


\begin{thebibliography}{99}
\bibitem{[N1]} P. A. M. Dirac, \textit{Lectures in Quantum Mechanics},
Benjamin, New York,1964.

A. Hanson, T. Regge and C. Teitelboim, \textit{Constrained Hamiltonian
Systems}, Acc. Naz. dei Lincei, Roma, 1976.

\bibitem{[N2]} P. A. M. Dirac, Rev. Mod. Phys. \textbf{21}, 392 (1949).

\bibitem{[N3]} Stanley J. Brodsky, Hans-Christian Pauli and Stephen S.
Pinsky, Phys. Rep. \textbf{301}, 299 (1988).

P. P. Srivastava, Nuovo Cimento \textbf{A107 }, 549 (1994).

\bibitem{[N4]} D. Biatti and L. Susskind, Phys. Lett. \textbf{B425}, 351
(1998).

\bibitem{[N5]} G. L Lapage and S. J. Brodsky. Phys. Rev. \textbf{D22}, 2157
(1980).

A. Bassetto, M. Dalbosco and R. Soldati, Phys. Rev. \textbf{D36}, 3138
(1987).

A. Bassetto, G. Heinrich, Z. Kunszt and W. Volgelsang, Phys. Rev. \textbf{D58%
}, 94020 (1998)

\bibitem{[N6]} E. Tomboulis, Phys. Rev. \textbf{D8}, 2736 (1971).

\bibitem{[N7]} G. McKeon, Can. J. Phys. \textbf{64}, 549 (1986).

\bibitem{[N8]} M. Marara, R. Soldati and G. McCartor. IAP Conf. Proc.
\textbf{494}, 284 (199).

\bibitem{[N9]} R. A. Neville and F. Rohrlich, Phys. Rev. \textbf{D3},
1692(1971).

\bibitem{[N10]} R. Casana, B. M. Pimentel and G. E. R. Zambrano, \emph{$%
SQED_4$ and $QED_4$ on the null-plane}, arXiv:0803.2677 [hep-th].

\bibitem{[N11]} F. Rohrlich, Acta Phys. Austriaca, Suppl. VIII, 277(1971).

R. A. Neville and F. Rohrlich, Nuovo Cimento \textbf{A1 }, 625 (1971).

\bibitem{[N13]} R. Benguria, P. Cordero and C. Teitelboim, Nucl. Phys.
\textbf{B122}, 61 (1976).

\bibitem{[N12]} P. J. Steinhardt, Ann. Phys \textbf{128}, 425 (1980).
\end{thebibliography}
\end{document}